\documentclass{PoS}
\pdfoutput=1

\usepackage{amsmath}
\usepackage{amssymb}

\title{Inclusive three jet production at the LHC at 7 and 13 TeV collision energies}

\ShortTitle{Inclusive 3-jet production}

\author{\speaker{G. Chachamis}\\
        Instituto de F{\' \i}sica Te{\' o}rica UAM/CSIC, Nicol{\'a}s Cabrera 15\\
        \& Universidad Aut{\' o}noma de Madrid, E-28049 Madrid, Spain.
        E-mail: \email{chachamis@gmail.com}}

\author{F. Caporale\\
        Instituto de F{\' \i}sica Te{\' o}rica UAM/CSIC, Nicol{\'a}s Cabrera 15\\
        \& Universidad Aut{\' o}noma de Madrid, E-28049 Madrid, Spain.
        E-mail: \email{francesco.caporale@uam.es}}
        
\author{F.~G. Celiberto\\
        Dipartimento di Fisica, Universit{\`a} della Calabria \&\\
        Istituto Nazionale di Fisica Nucleare, Gruppo Collegato di Cosenza,\\
        I-87036 Arcavacata di Rende, Cosenza, Italy.
        E-mail: \email{francescogiovanni.celiberto@fis.unical.it}}

\author{D. Gordo G{\' o}mez\\
 Instituto de F{\' \i}sica Te{\' o}rica UAM/CSIC, Nicol{\'a}s Cabrera 15\\
        \& Universidad Aut{\' o}noma de Madrid, E-28049 Madrid, Spain.
        E-mail: \email{david.gordo@csic.es}}

\author{A. Sabio Vera\\
 Instituto de F{\' \i}sica Te{\' o}rica UAM/CSIC, Nicol{\'a}s Cabrera 15\\
        \& Universidad Aut{\' o}noma de Madrid, E-28049 Madrid, Spain.
        E-mail: \email{a.sabio.vera@gmail.com}}

\abstract{We discuss briefly a recent study of new observables in LHC inclusive events with three tagged jets.
One jet is in the forward direction, the second is in the backward direction and well-separated in rapidity
from the first, whereas, the third tagged jet is to be found in more central regions of  the detector. 
Taking into consideration that non-tagged mini-jet emissions are allowed and that they may be accounted for by the BFKL gluon Green function, we project the cross sections on azimuthal-angle components and define
suitable ratios based on these projections which can provide several distinct tests of the BFKL dynamics.}

\FullConference{XXIV International Workshop on Deep-Inelastic Scattering and Related Subjects\\
		11-15 April, 2016\\
		DESY Hamburg, Germany}

\begin{document}

\section{Introduction}

At hadronic colliders, when jets are produced at large relative rapidities the Balitsky-Fadin-Kuraev-Lipatov (BFKL) framework in the leading logarithmic (LL)~\cite{Lipatov:1985uk,Balitsky:1978ic,Kuraev:1977fs,Kuraev:1976ge,Lipatov:1976zz,Fadin:1975cb} and next-to-leading logarithmic (NLL) approximation~\cite{Fadin:1998py,Ciafaloni:1998gs} is applicable.  
Mueller-Navelet jets~\cite{Mueller:1986ey}, in particular 
ratios of projections on azimuthal-angle observables~${\cal R}^m_n = \langle \cos{(m \, \phi)} \rangle / \langle \cos{(n \, \phi)} \rangle$, for the azimuthal-angle  formed by the two tagged jets, $\phi$, is an important example
in which comparison of different NLL predictions with LHC experimental data has  been quite successful
~\cite{DelDuca:1993mn,Stirling:1994he,Orr:1997im,Kwiecinski:2001nh,Angioni:2011wj,Caporale:2013uva, Caporale:2013sc,Marquet:2007xx,Colferai:2010wu,Ducloue:2013wmi,Ducloue:2014koa,Mueller:2015ael,Vera:2006un,Vera:2007kn,Ducloue:2013bva,Caporale:2014gpa,Caporale:2014blm,Celiberto:2015dgl,Celiberto:2016hae,Celiberto:2016ygs}.

Recently, we proposed new observables for processes at the LHC that may be considered as a generalisation
of the Mueller-Navelet jets. These processes are inclusive three-jet~\cite{Caporale:2015vya,Caporale:2016soq}
and four-jet production~\cite{Caporale:2015int,Caporale:2016xku}. Here we will solely focus on the three-jet
observables.
These are based on the recently defined ratios~\cite{Caporale:2015vya}
\begin{eqnarray}
{\cal R}^{M N}_{P Q} =\frac{ \langle \cos{(M \, \phi_1)} \cos{(N \, \phi_2)} \rangle}{\langle \cos{(P \, \phi_1)} \cos{(Q \, \phi_2)} \rangle} \, , 
\label{Rmnpq}
\end{eqnarray}
where $\phi_1$ and $\phi_2$ are, respectively, the azimuthal angle difference between the first and the second (central) jet and between this one and the third jet.
The ``observables" ${\cal R}^{M N}_{P Q}$ in Eq.~(\ref{Rmnpq}) are defined at partonic level and cannot be 
readily compared to experimental data. For that reason, we report here on the hadronic level
observables $R^{M N}_{P Q}$~\cite{Caporale:2016soq}. 
This will allow for a comparison of our predictions
 with forthcoming analyses of the LHC experimental data. 
 Within the collinear factorization scheme we produce the two utmost in rapidity jets and we associate to each 
 one of them a forward ``jet vertex''~\cite{Caporale:2012:IF} before we connect these vertices with the 
 central jet via two BFKL gluon Green functions. Finally we convolute the partonic differential
 cross-section with collinear parton distribution functions
 and we integrate over the momenta of all produced jets,
 using  LHC experimental cuts, to compute the ratios $R^{M N}_{P Q}$. We fix the rapidity of the central jet to lie in the middle of the two utmost tagged jets.

\section{Hadronic inclusive three-jet production in multi-Regge kinematics}
Let us recapitulate some of the notation used in~\cite{Caporale:2015vya,Caporale:2016soq}.
If the transverse momenta of the utmost jets are $\vec{k}_{A,B}$ and their rapidity distance $Y$ is large
while the central jet has transverse momentum $\vec{k}_J$ and mini-jet activity is allowed between the
three tagged jets, the studied process is
\begin{eqnarray}
\label{process}
{\rm proton }(p_1) + {\rm proton} (p_2) \to  
{\rm jet}(k_A) + {\rm jet}(k_J) + {\rm jet}(k_B)  + {\rm minijets}\;.
\end{eqnarray}

The projection on azimuthal-angle components can give
the mean value (with $M,N$ being positive integers)
\begin{eqnarray}
\label{Cmn}
 {\cal C}_{MN} \, = \,
 \langle \cos{\left(M \left( \theta_A - \theta_J - \pi\right)\right)}  
 \cos{\left(N \left( \theta_J - \theta_B - \pi\right)\right)}
 \rangle && \\
 &&\hspace{-9cm} = \frac{\int_0^{2 \pi} d \theta_A d \theta_B d \theta_J \cos{\left(M \left( \theta_A - \theta_J - \pi\right)\right)}  \cos{\left(N \left( \theta_J - \theta_B - \pi\right)\right)}
 d\sigma^{3-{\rm jet}} }{\int_0^{2 \pi} d \theta_A d \theta_B d \theta_J 
 d\sigma^{3-{\rm jet}} },\nonumber
\end{eqnarray}
where we define
the two relative azimuthal angles between each external jet
and the central one as
$\Delta\theta_{\widehat{AJ}} = \theta_A - \theta_J - \pi$ and 
$\Delta\theta_{\widehat{JB}} = \theta_J - \theta_B - \pi$ respectively
and  $d\sigma^{3-{\rm jet}}$ is defined in ~\cite{Caporale:2015vya}.

As we mentioned previously,
our main target is to provide theoretical estimates that can be compared against
current and future experimental data, therefore, we introduce kinematical cuts already 
in use at the LHC. We integrate ${\cal C}_{M,N}$ over the momenta of the tagged jets in the form
\begin{align}
\label{Cmn_int}
 C_{MN} =
 \int_{Y_A^{\rm min}}^{Y_A^{\rm max}} \hspace{-0.25cm} dY_A
 \int_{Y_B^{\rm min}}^{Y_B^{\rm max}} \hspace{-0.25cm} dY_B
 \int_{k_A^{\rm min}}^{k_A^{\rm max}} \hspace{-0.25cm} dk_A
 \int_{k_B^{\rm min}}^{k_B^{\rm max}} \hspace{-0.25cm} dk_B
 \int_{k_J^{\rm min}}^{k_J^{\rm max}} \hspace{-0.25cm} dk_J
 \delta\left(Y_A - Y_B - Y\right) {\cal C}_{MN},
\end{align}
where the rapidities of the utmost jet rapidities take values in the
range  $Y_{A,\, B}^{\rm min}  = -4.7$  and 
$Y_{A,\, B}^{\rm max}  = 4.7$ while  their difference 
$Y \equiv Y_A - Y_B$ is kept fixed at definite values in the range $5 < Y < 9$.
We compute for two different 
center-of-mass energies, $\sqrt s = 7$ and $\sqrt s = 13$ TeV, and we use both a symmetric and an asymmetric
cut~\cite{Ducloue:2013wmi,Celiberto:2015dgl}:
\begin{enumerate}
\item $k_A^{\rm min} = 35$ GeV, $k_B^{\rm min} = 35$ GeV, $k_A^{\rm max} = k_B^{\rm max}  = 60$ GeV
(symmetric); \,
\item $k_A^{\rm min} = 35$ GeV, $k_B^{\rm min} = 50$ GeV,  $k_A^{\rm max} = k_B^{\rm max}  = 60$ GeV
(asymmetric).
\end{enumerate}

\begin{figure}[h]
\centering
   \includegraphics[scale=0.35]{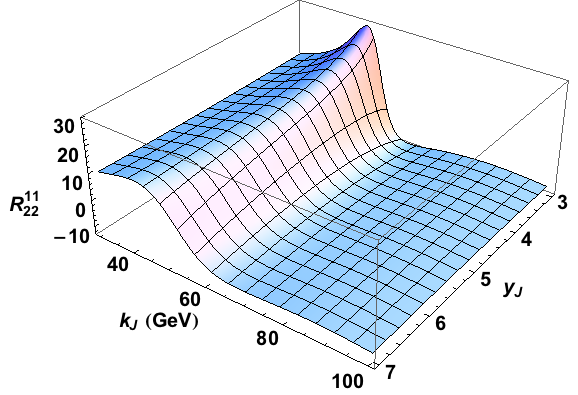}
\caption{\small 3D plot of the partonic ${\cal R}^{11}_{22}$ as a function 
of the momentum $k_J$ and the rapidity $y_J$ of the 
central jet for $k_A = 40$ GeV, $k_B = 50$ GeV and $\Delta Y_{A,B}=10$.} 
\label{3d}
\end{figure}

Seeking the best possible perturbative 
stability  in our 
results (see~\cite{Caporale:2013uva} for a related discussion). 
we remove the  
zeroth conformal spin contribution of the BFKL kernel.
by introducing the ratios
\begin{eqnarray}
\label{RPQMN}
R_{PQ}^{MN} \, = \, \frac{C_{MN}}{C_{PQ}}, \,\,\,\,\,\,\,\, M, N, P, \,Q > 0,
\label{RmnqpNew}
\end{eqnarray}
which are free from any $n=0$ dependence. Thus, 
we can study the ratios  $R_{PQ}^{MN}(Y)$ in Eq.~(\ref{RmnqpNew}) as functions of the 
rapidity difference $Y$ between the utmost jets 
for a set of typical values of $M, N, P, Q$.
The momentum of the central jet is permitted to take values in three different
domains: $[20\, \mathrm{GeV} < k_J < 35\, \mathrm{GeV}]$  (smaller that $k_A$, $k_B$),
$[35 \,\mathrm{GeV} < k_J < 60\, \mathrm{GeV}]$ (similar to $k_A$, $k_B$) and
$[60\, \mathrm{GeV} < k_J < 120\, \mathrm{GeV}]$ (larger than $k_A$, $k_B$). This
 allows us to see how the ratio
$R_{PQ}^{MN}(Y)$ changes behaviour depending on the relative size of the
three jets, see Fig.~\ref{3d} for the behaviour of ${\cal R}^{11}_{22}$.

\begin{figure}[h]
\vspace{-.3cm}
\centering

   \hspace{-.6cm}
   \includegraphics[scale=0.38]{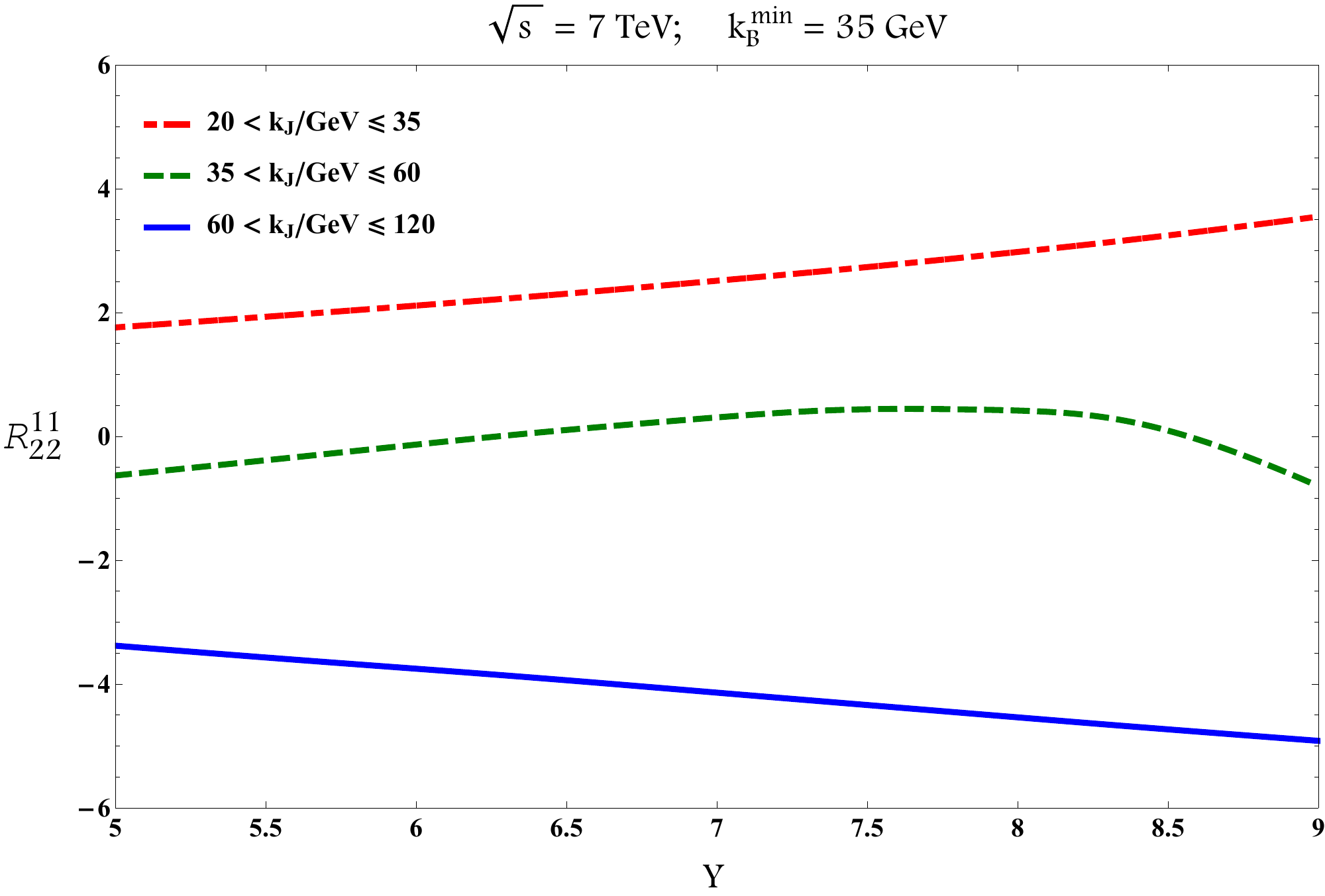}
   \includegraphics[scale=0.38]{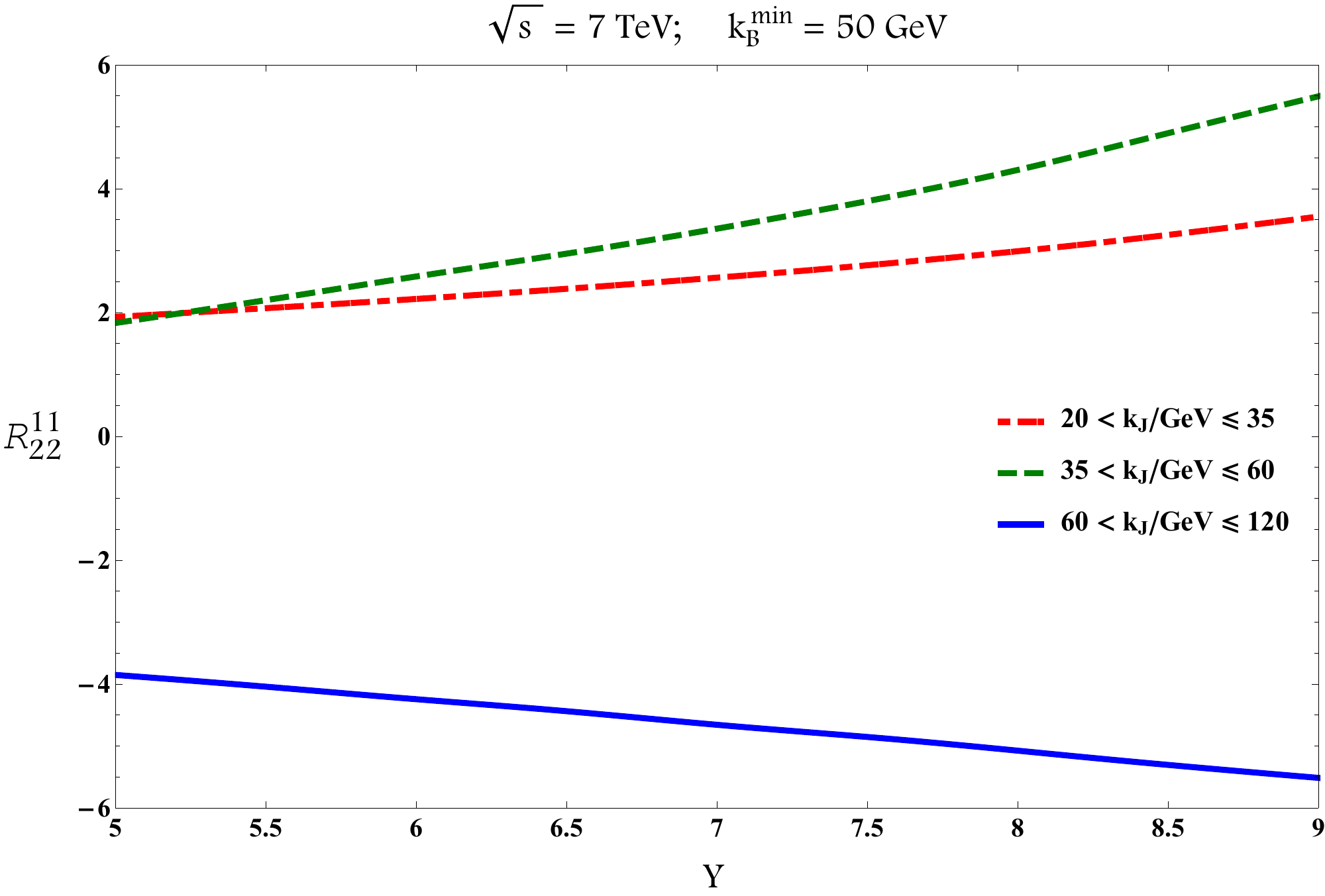}
   \vspace{.5cm}

   \hspace{-.6cm}
   \includegraphics[scale=0.38]{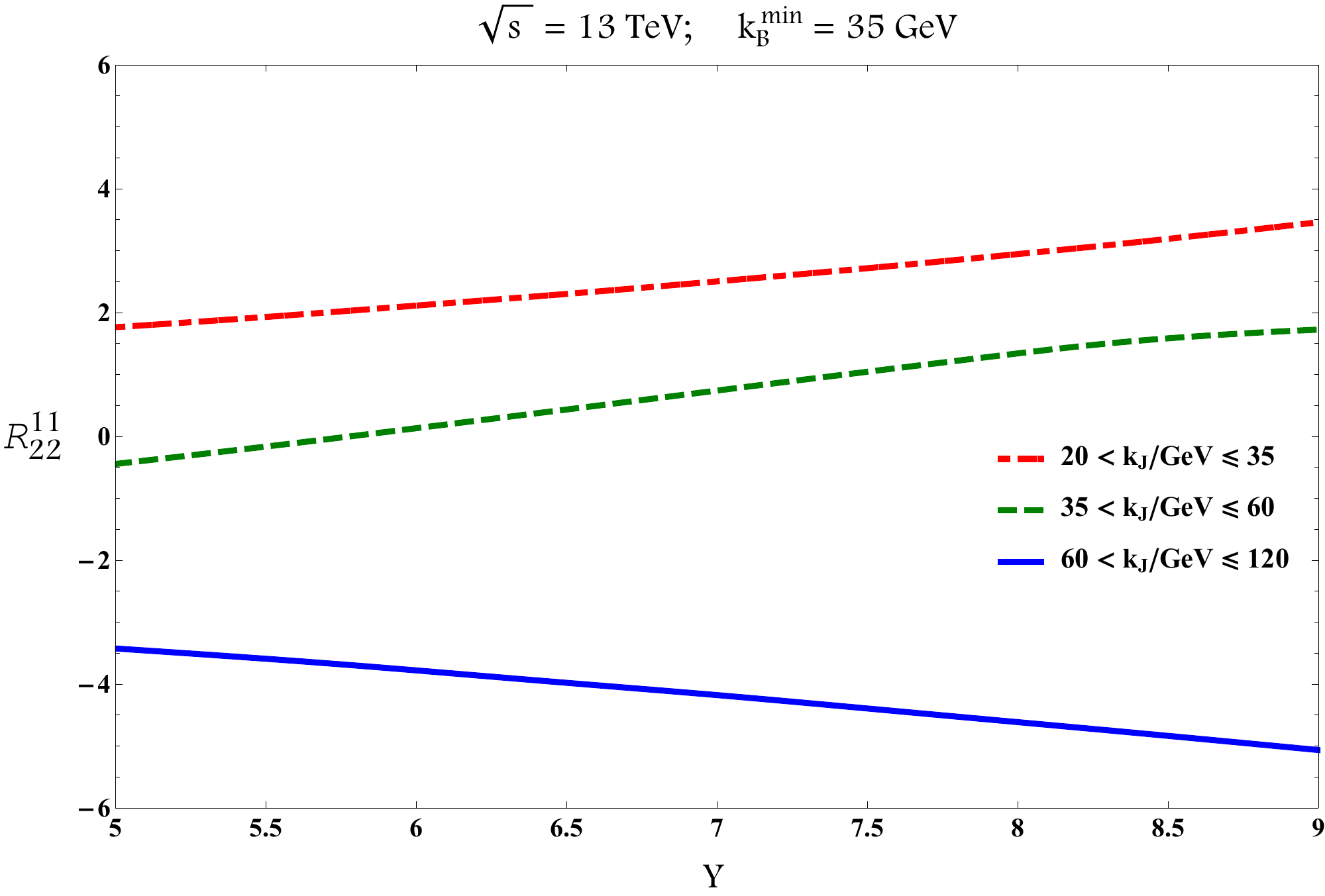}
   \includegraphics[scale=0.38]{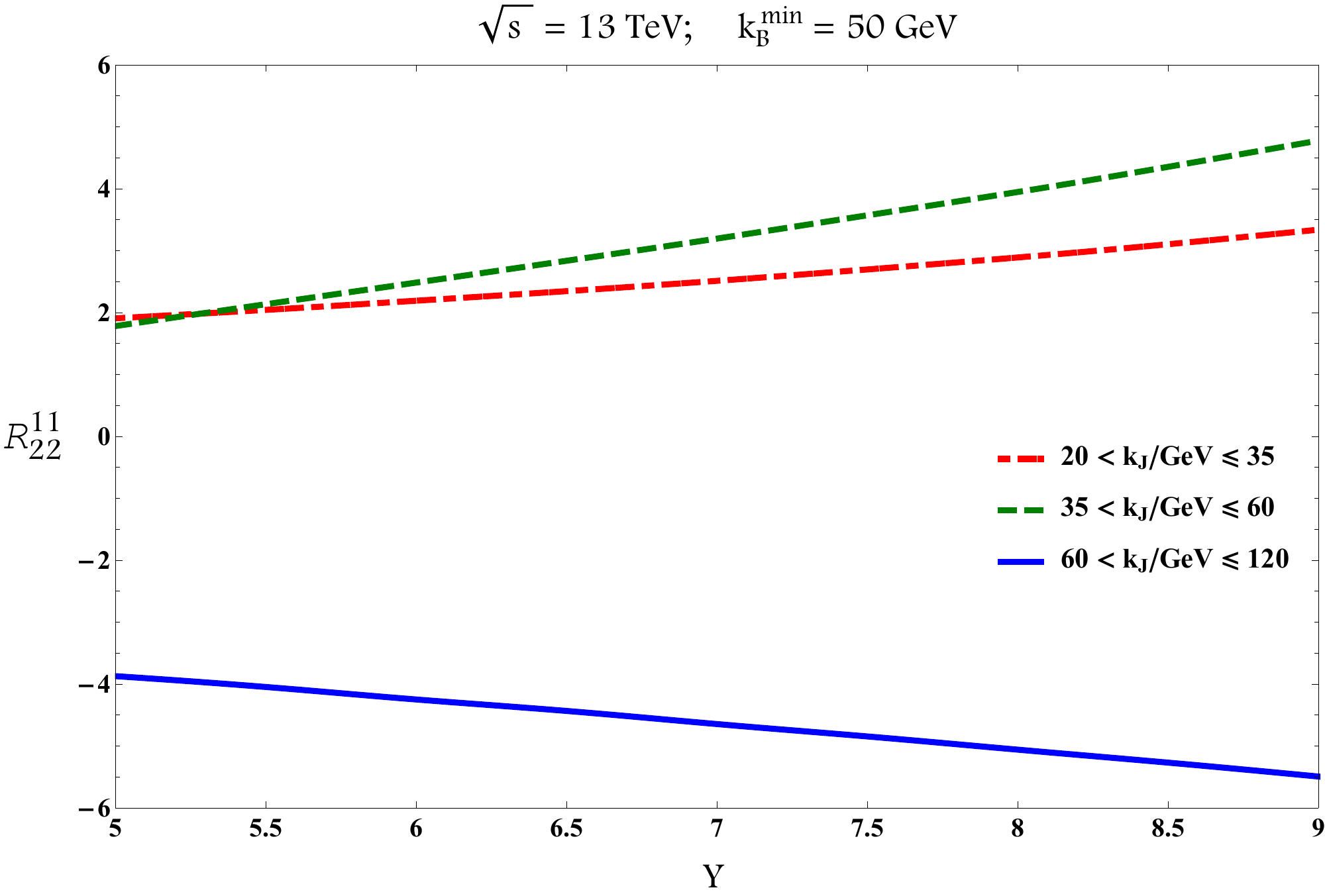}

\caption{\small $Y$-dependence of 
$R^{11}_{22}$ for $\sqrt s = 7$ TeV (top) and  $\sqrt s = 13$ TeV (bottom).
$k_B^{\rm min} = 35$ GeV (left column) and $k_B^{\rm min} = 50$ GeV (right column).} 
\label{res}
\end{figure}
 
In total, we have computed the results
for six different ratios in~\cite{Caporale:2016soq}, 
here we are only showing for $R^{11}_{22}$ in Fig.~\ref{res}.
Generally, the dependence of the different observables on the rapidity
difference between $k_A$ and $k_B$ is rather smooth. 
The slope of the three curves, in absolute values, depends
on the particular observable. 
Another interesting observation is that there are ratios for which
changing from the symmetric to the asymmetric cut makes no real difference
and other ratios for which the picture changes drastically.
The most important thing to note though would be that, in general, 
for most of the ratios there are no
large scale changes when we go up to $\sqrt{s} = 13$ TeV from $\sqrt{s} = 7$ TeV.
This is indeed significant 
as it seemingly suggests that we reach a sort of asymptotic regime 
for the kinematical configurations used in our study. It also indicates 
that our ratios
are indeed mostly insensitive  to effects
stemming from outside the BFKL dynamics and which 
normally cannot be detached ({\it e.g.} influence from the PDFs).

\section{Summary \& Outlook}

We have discussed a first phenomenological study 
at hadronic level of azimuthal-angle dependent observables in
inclusive three-jet production at the LHC within the BFKL resummation program.
The study was focused on how the ratios
$R_{M N}^{P Q}$,
behave when we change the rapidity
difference $Y$ between the utmost jets from 5 to 9 units.
In general, a smooth functional dependence of
the ratios on $Y$ is seen. A major observation is that
these observables do not have a significantly different behaviour after
raising the colliding energy from 7 to 13 TeV which make us confident  that
they highlight the most important features of the tagged jets azimuthal behavior
within the BFKL dynamics.
It will be  important to see whether  
fixed order calculations and studies with the BFKL inspired 
Monte Carlo  {\bf BFKLex}~\cite{Chachamis:2011rw,Chachamis:2011nz,Chachamis:2012fk,
Chachamis:2012qw,Caporale:2013bva,Chachamis:2015zzp,Chachamis:2015ico}. 
give similar predictions. Estimates from the usual all-purpose collinear
Monte Carlo tools are also needed to have a full picture from the theoretical side.
In conclusion, 
only an experimental analysis for these observables using existing and future LHC data
 will show us whether we
can use these observables as a new probe for the BFKL dynamics and if so,
it will help toward the direction of assessing the window of applicability 
of the BFKL resummation program.

\begin{flushleft}
\vspace{-.3cm}
{\bf \large Acknowledgements}
\end{flushleft}
\vspace{-.3cm}
GC acknowledges support from the MICINN, Spain, 
under contract FPA2013-44773-P. 
ASV acknowledges support from Spanish Government 
(MICINN (FPA2010-17747,FPA2012-32828)) and, together with FC and FGC, 
to the Spanish MINECO Centro de Excelencia Severo Ochoa Programme 
(SEV-2012-0249). FGC thanks the Instituto de F{\'\i}sica Te{\'o}rica 
(IFT UAM-CSIC) in Madrid for warm hospitality.


\begin{thebibliography}{99}

\bibitem{Lipatov:1985uk}
  L.~N.~Lipatov,
  Sov.\ Phys.\ JETP {\bf 63} (1986) 904
   [Zh.\ Eksp.\ Teor.\ Fiz.\  {\bf 90} (1986) 1536].
  
\bibitem{Balitsky:1978ic}
  I.~I.~Balitsky and L.~N.~Lipatov,
  Sov.\ J.\ Nucl.\ Phys.\  {\bf 28} (1978) 822
   [Yad.\ Fiz.\  {\bf 28} (1978) 1597].
  
\bibitem{Kuraev:1977fs}
  E.~A.~Kuraev, L.~N.~Lipatov and V.~S.~Fadin,
  Sov.\ Phys.\ JETP {\bf 45} (1977) 199
   [Zh.\ Eksp.\ Teor.\ Fiz.\  {\bf 72} (1977) 377].
  
\bibitem{Kuraev:1976ge}
  E.~A.~Kuraev, L.~N.~Lipatov and V.~S.~Fadin,
  Sov.\ Phys.\ JETP {\bf 44} (1976) 443
   [Zh.\ Eksp.\ Teor.\ Fiz.\  {\bf 71} (1976) 840]
   [Erratum-ibid.\  {\bf 45} (1977) 199].
  
\bibitem{Lipatov:1976zz}
  L.~N.~Lipatov,
  Sov.\ J.\ Nucl.\ Phys.\  {\bf 23} (1976) 338
   [Yad.\ Fiz.\  {\bf 23} (1976) 642].
  
\bibitem{Fadin:1975cb}
  V.~S.~Fadin, E.~A.~Kuraev and L.~N.~Lipatov,
  Phys.\ Lett.\ B {\bf 60} (1975) 50.
  
\bibitem{Fadin:1998py}
  V.~S.~Fadin and L.~N.~Lipatov,
  Phys.\ Lett.\ B {\bf 429} (1998) 127
  [hep-ph/9802290].
  
\bibitem{Ciafaloni:1998gs}
  M.~Ciafaloni and G.~Camici,
  Phys.\ Lett.\ B {\bf 430} (1998) 349
  [hep-ph/9803389].

\bibitem{Mueller:1986ey}
  A.~H.~Mueller and H.~Navelet,
  Nucl.\ Phys.\ B {\bf 282} (1987) 727.


\bibitem{DelDuca:1993mn}
  V.~Del Duca and C.~R.~Schmidt,
  Phys.\ Rev.\ D {\bf 49} (1994) 4510
  [hep-ph/9311290].
  
\bibitem{Stirling:1994he}
  W.~J.~Stirling,
  Nucl.\ Phys.\ B {\bf 423} (1994) 56
  [hep-ph/9401266].
  
\bibitem{Orr:1997im}
  L.~H.~Orr and W.~J.~Stirling,
  Phys.\ Rev.\ D {\bf 56} (1997) 5875
  [hep-ph/9706529].
  
\bibitem{Kwiecinski:2001nh}
  J.~Kwiecinski, A.~D.~Martin, L.~Motyka and J.~Outhwaite,
  Phys.\ Lett.\ B {\bf 514} (2001) 355
  [hep-ph/0105039].
  
\bibitem{Angioni:2011wj} 
  M.~Angioni, G.~Chachamis, J.~D.~Madrigal and A.~Sabio Vera,
  Phys.\ Rev.\ Lett.\  {\bf 107}, 191601 (2011)
  doi:10.1103/PhysRevLett.107.191601
  [arXiv:1106.6172 [hep-th]].

  
\bibitem{Caporale:2013uva}
  F.~Caporale, B.~Murdaca, A.~Sabio Vera and C.~Salas,
  Nucl.\ Phys.\ B {\bf 875} (2013) 134
  [arXiv:1305.4620 [hep-ph]].

\bibitem{Caporale:2013sc}
  F.~Caporale, D.~Y.~Ivanov, B.~Murdaca and A.~Papa,
  Nucl.\ Phys.\ B {\bf 877} (2013) 73
  [arXiv:1211.7225 [hep-ph]].

\bibitem{Marquet:2007xx} 
  C.~Marquet and C.~Royon,
  Phys.\ Rev.\ D {\bf 79}, 034028 (2009)
  doi:10.1103/PhysRevD.79.034028
  [arXiv:0704.3409 [hep-ph]].
  
\bibitem{Colferai:2010wu} 
  D.~Colferai, F.~Schwennsen, L.~Szymanowski and S.~Wallon,
  JHEP {\bf 1012}, 026 (2010)
  doi:10.1007/JHEP12(2010)026
  [arXiv:1002.1365 [hep-ph]].
  
\bibitem{Ducloue:2013wmi} 
  B.~Ducloue, L.~Szymanowski and S.~Wallon,
  JHEP {\bf 1305}, 096 (2013)
  doi:10.1007/JHEP05(2013)096
  [arXiv:1302.7012 [hep-ph]].

\bibitem{Ducloue:2014koa} 
  B.~Ducloue, L.~Szymanowski and S.~Wallon,
  Phys.\ Lett.\ B {\bf 738}, 311 (2014)
  doi:10.1016/j.physletb.2014.09.025
  [arXiv:1407.6593 [hep-ph]].

\bibitem{Mueller:2015ael} 
  A.~H.~Mueller, L.~Szymanowski, S.~Wallon, B.~W.~Xiao and F.~Yuan,
  JHEP {\bf 1603}, 096 (2016)
  doi:10.1007/JHEP03(2016)096
  [arXiv:1512.07127 [hep-ph]].

\bibitem{Vera:2006un}
  A.~Sabio Vera,
  Nucl.\ Phys.\ B {\bf 746} (2006) 1
  [hep-ph/0602250].
  
\bibitem{Vera:2007kn}
  A.~Sabio Vera and F.~Schwennsen,
  Nucl.\ Phys.\ B {\bf 776} (2007) 170
  [hep-ph/0702158 [HEP-PH]].

     
\bibitem{Ducloue:2013bva}
  B.~Ducloue, L.~Szymanowski and S.~Wallon,
  Phys.\ Rev.\ Lett.\  {\bf 112} (2014) 082003
  [arXiv:1309.3229 [hep-ph]].

\bibitem{Caporale:2014gpa}
  F.~Caporale, D.~Y.~Ivanov, B.~Murdaca and A.~Papa,
  Eur.\ Phys.\ J.\ C {\bf 74} (2014) 3084
  [arXiv:1407.8431 [hep-ph]].

\bibitem{Caporale:2014blm}
  F.~Caporale, D.~Y.~Ivanov, B.~Murdaca and A.~Papa,
Phys.\ Rev.\ D{\bf 91} (2015) 11, 114009 
  [arXiv:1504.06471 [hep-ph]].

\bibitem{Celiberto:2015dgl}
 F.~G. Celiberto, D.~Yu. Ivanov, B.~Murdaca and A.~Papa,
 Eur.\ Phys.\ J.\ C {\bf 75} (2015) 292 
 [arXiv:1504.08233 [hep-ph]].

\bibitem{Celiberto:2016hae} 
  F.~G.~Celiberto, D.~Y.~Ivanov, B.~Murdaca and A.~Papa,
  arXiv:1604.08013 [hep-ph].

\bibitem{Celiberto:2016ygs} 
  F.~G.~Celiberto, D.~Y.~Ivanov, B.~Murdaca and A.~Papa,
  Eur.\ Phys.\ J.\ C {\bf 76}, no. 4, 224 (2016)
  doi:10.1140/epjc/s10052-016-4053-5
  [arXiv:1601.07847 [hep-ph]].

\bibitem{Caporale:2015vya} 
  F.~Caporale, G.~Chachamis, B.~Murdaca and A.~Sabio~Vera,
  Phys.\ Rev.\ Lett.\  {\bf 116}, no. 1, 012001 (2016)
  doi:10.1103/PhysRevLett.116.012001
  [arXiv:1508.07711 [hep-ph]].
    
\bibitem{Caporale:2016soq} 
  F.~Caporale, F.~G.~Celiberto, G.~Chachamis, D.~G.~Gomez and A.~Sabio~Vera,
  arXiv:1603.07785 [hep-ph].
    
\bibitem{Caporale:2015int} 
  F.~Caporale, F.~G.~Celiberto, G.~Chachamis and A.~Sabio~Vera,
  Eur.\ Phys.\ J.\ C {\bf 76}, no. 3, 165 (2016)
  doi:10.1140/epjc/s10052-016-3963-6
  [arXiv:1512.03364 [hep-ph]].
  
\bibitem{Caporale:2016xku} 
  F.~Caporale, F.~G.~Celiberto, G.~Chachamis, D.~G.~Gomez and A.~Sabio~Vera,
  arXiv:1606.00574 [hep-ph].
  

\bibitem{Caporale:2012:IF}
F. Caporale, D. Yu. Ivanov, B. Murdaca, A.Papa, A.Perri, JHEP {\bf 1202} (2012) 101; 
 [arXiv:1212.0487 [hep-ph]].
  
   
 
\bibitem{Chachamis:2011rw}
  G.~Chachamis, M.~Deak, A.~Sabio Vera and P.~Stephens,
 Nucl.\ Phys.\ B {\bf 849} (2011) 28
  [arXiv:1102.1890 [hep-ph]].

\bibitem{Chachamis:2011nz}
  G.~Chachamis and A.~Sabio Vera,
  Phys.\ Lett.\ B {\bf 709} (2012) 301
  [arXiv:1112.4162 [hep-th]].

\bibitem{Chachamis:2012fk}
  G.~Chachamis and A.~Sabio Vera,
  Phys.\ Lett.\ B {\bf 717} (2012) 458
  [arXiv:1206.3140 [hep-th]].

\bibitem{Chachamis:2012qw}
  G.~Chachamis, A.~Sabio Vera and C.~Salas,
  Phys.\ Rev.\ D {\bf 87} (2013) 1,  016007
  [arXiv:1211.6332 [hep-ph]].

\bibitem{Caporale:2013bva}
  F.~Caporale, G.~Chachamis, J.~D.~Madrigal, B.~Murdaca and A.~Sabio Vera,
  Phys.\ Lett.\ B {\bf 724} (2013) 127
  [arXiv:1305.1474 [hep-th]].

\bibitem{Chachamis:2015zzp} 
  G.~Chachamis and A.~Sabio Vera,
   arXiv:1511.03548 [hep-ph].
 
\bibitem{Chachamis:2015ico}
  G.~Chachamis and A.~Sabio~Vera,
  JHEP {\bf 1602} (2016) 064
  doi:10.1007/JHEP02(2016)064
  [arXiv:1512.03603 [hep-ph]].

\end{thebibliography}
\end{document}